# Towards a Blockchain and Opportunistic Edge Driven Metaverse of Everything

Paula Fraga-Lamas*, Sérgio Ivan Lopes†‡§¶, Tiago M. Fernández-Caramés*
*Dpt. of Computer Engineering, Centro de investigación CITIC,
Faculty of Computer Science, Universidade da Coruña, 15071, A Coruña, Spain
†ADiT-Lab, Instituto Politécnico de Viana do Castelo, 4900-348 Viana do Castelo, Portugal
‡CiTin—Centro de Interface Tecnológico Industrial, 4970-786 Arcos de Valdevez, Portugal
§IT—Instituto de Telecomunicações, Campus Universitário de Santiago, 3810-193 Aveiro, Portugal
Email: paula.fraga@udc.es, sil@estg.ipvc.pt, tiago.fernandez@udc.es

*Abstract*—Decentralized Metaverses, built on Web 3.0 and Web 4.0 technologies, have attracted significant attention across various fields. This innovation leverages blockchain, Decentralized Autonomous Organizations (DAOs), Extended Reality (XR) and advanced technologies to create immersive and interconnected digital environments that mirror the real world. This article delves into the Metaverse of Everything (MoE), a platform that fuses the Metaverse concept with the Internet of Everything (IoE), an advanced version of the Internet of Things (IoT) that connects not only physical devices but also people, data and processes within a networked environment. Thus, the MoE integrates generated data and virtual entities, creating an extensive network of interconnected components. This article seeks to advance current MoE, examining decentralization and the application of Opportunistic Edge Computing (OEC) for interactions with surrounding IoT devices and IoE entities. Moreover, it outlines the main challenges to guide researchers and businesses towards building a future cyber-resilient opportunistic MoE.

*Index Terms*—Metaverse of Everything; IoT; Blockchain; Opportunistic Edge Computing; Digital Twins; Extended Reality; Web3; Web 4.0; Future Internet.

## I. Introduction

The Metaverse can be defined as a network of real-time rendered 3D virtual worlds delivered through Extended Reality (XR) devices like Augmented and Mixed Reality (AR/MR) smart glasses or Virtual Reality (VR) headsets [1], [2]. The term "Metaverse" originates from Neal Stephenson's 1992 science fiction novel Snow Crash. However it gained substantial attention in late 2021 following Mark Zuckerberg's announcement of Facebook's rebranding to Meta, a new company dedicated to developing next-generation social networking solutions. Since then, the Metaverse became a popular buzzword among companies, each interpreting and promoting it according to their own perspectives and technological capabilities.

This work has been funded by grants TED2021-129433A-C22 (HELENE) and PID2020-118857RA-100 (ORBALLO) funded by MCIN/AEI/10.13039/501100011033 and the European Union NextGenerationEU/PRTR, and by Project "TEXPCT: Pacto de inovacção para a digitalização do STV", funded under the Recovery and Resilience Plan – 02-C05-i01.01-2022.PC644915249-00000025, under the scope of the project "WP 2.2 (PPS#8): Digital Twin Platform for Textile Manufacturing".

Few articles in the literature examine decentralized metaverses that rely on blockchain [3], with the majority focusing on cybersecurity and Decentralized Autonomous Organizations (DAOs). In particular, in [4] the concept of Metaverse of Everything (MoE) is defined as a concept that merges the Metaverse with the Internet of Everything (IoE), an extension of the Internet of Things (IoT) paradigm. In [4] the authors outline the main opportunities and challenges and provide a thorough SWOT analysis. Nevertheless, the article does not address the need to incorporate Opportunistic Edge Computing (OEC) communications [2] to build hybrid worlds that allow for interacting seamlessly with both virtual and physical entities.

In contrast, the motivation behind this article is to explore how blockchain and opportunistic communications can enhance the MoE by addressing current limitations related to cybersecurity and latency in the communications of an edge-cloud computing continuum. To the knowledge of the authors, this is the first article that proposes a global and comprehensive vision of how to confront the main MoE issues to guide researchers and managers on future developments. This article also aims to shed light on the integration of XR, blockchain, IoE and opportunistic communications, providing valuable insights into the field.

The rest of this paper is organized as follows. Section II provides a conceptualization the Metaverse of Everything (MoE) and an overview of its key concepts, Section III proposes a communications architecture for the MoE. In Section IV the main challenges of the MoE are outlined. Finally, Section V is devoted to conclusions.

## II. Basic concepts

### A. The Metaverse

The Metaverse is envisioned as an extension of the physical world where Metaverse users (Meta-users) can interact with a computer-generated environment and with other Meta-users [1], [5]–[7], enabling interaction across diverse virtual or hybrid spaces (i.e., a space that is partially real and partially virtual). Therefore, it is a network of shared interconnected spaces that exist alongside the physical world. Such a concept is set to drive the future Internet and to influence all sectors of the economy, like entertainment, education, industry (with the so-called industrial metaverses [2]), finance, and even governance [8]. By leveraging advancements in ubiquitous computing and sophisticated technologies, the Metaverse provides a versatile framework for novel applications.

From an engineering standpoint, metaverses exhibit typical characteristics of Cyber-Physical Systems (CPSs) [9] or even Cyber-Physical Social Systems (CPSSs) [10], which specially refers to three worlds (physical, virtual, and mental worlds). Despite its potential, the Metaverse lacks a standardized definition, and there is even an ongoing debate about the core technologies it encompasses (e.g., it is not clear whether it is indispensable to use ultra-low latency communication networks like 6G). Nonetheless, XR technologies are undeniably central to its development thanks to their ability to offer novel and immersive ways to interact with digital content. For instance, XR technologies like Augmented Reality (AR) enable overlaying digital information onto the physical world; Virtual Reality (VR) can create fully digital immersive environments; and Mixed Reality (MR) can combine real and virtual elements.

As the Metaverse continues to evolve, there is a growing need for decentralization to ensure user control, cybersecurity, and interoperability across different platforms [11]. In addition, IoT and Industrial IoT (IIoT) technologies are also crucial for bridging the physical and digital worlds since they provide real-time data through sensors and enable interactions with the environment through actuators [2].

B. Internet of Everything

The growing penetration of IoT technologies is reflected in sectors such as consumer electronics or CPSs for industrial automation [9]. The Internet of Everything (IoE) extends the IoT paradigm by connecting people, processes, data and things, forming a challenging integrated network of interconnected entities [12]. The added value of IoE lies in its ability to provide Machine-to-Machine (M2M), Person-to-Machine (P2M), and Person-to-Person connections [13]. For instance, Digital Twins leverage information provided by IoE entities by creating detailed digital models of physical objects, systems, or processes that are enriched with real-world data, thus allowing real-time simulations, analytics and optimizations, closing the loop control at the system level [4]. Thus, the integration of the IoE within the Metaverse creates the so-called Metaverse of Everything (MoE), which opens up a plethora of opportunities across various domains [4]. Table I shows some examples of IoE and MoE entities. Despite these opportunities, several challenges remain, including technical limitations, user adoption and ethical considerations [4], [7].

C. Web 3.0 and 4.0

Web 3.0, also known as Web3, is characterized by decentralization, user's full empowerment, and enhanced privacy [14]. Unlike its predecessor, Web 2.0, which is dominated by centralized platforms, Web 3.0 leverages Distributed Ledger Technologies (DLTs) like Blockchain and distributed networks to enable peer-to-peer interactions without central authorities [15]. DAOs can operate autonomously through smart contracts (self-executing contracts to automate transactions and interactions based on predefined conditions), eliminating the need for external intervention [15]. This new Web 3.0 paradigm emphasizes user control over personal data and digital assets (e.g., Non-Fungible Tokens (NFTs), dynamic NFTs (dNFTs), tokens), utilizes smart contracts for automated transactions, and supports decentralized applications (DApps) that run on distributed networks [4], [15]. Thus, by incorporating decentralized governance through DAOs, Web 3.0 aims to create a more open, secure, and user-centric Internet.

While Web 3.0 focuses on decentralization and openness, the so-called Web 4.0 is defined in an European Union initiative as a ground-breaking technological transition towards a world where everything is seamlessly interconnected [16].

D. The Metaverse of Everything

The MoE is an advanced conceptualization of the Metaverse that integrates and extends IoE into a unified immersive virtual environment. In the MoE, physical and digital worlds are seamlessly interconnected, with a comprehensive network of virtual entities, including people, objects, data and processes, all interacting within a persistent 3D virtual space. This environment leverages Web 3.0 technologies to create a decentralized and user-centric ecosystem where virtual representations of real-world assets and activities coexist. MoE is expected to revolutionize sectors, from entertainment to industry and work interactions [7].

E. Blockchain and Decentralized Metaverses

Decentralized metaverses like Decentraland or Voxels Metaverse are powered by Web 3.0 technologies, where virtual entities are stored, updated, shared and exchanged across local nodes. Decentralized intelligence circulates and aggregates among these nodes, producing integrated intelligence [10].

These platforms vary in their approach to governance, user control and interoperability [11], [17]. In the case of centralized metaverses, they are vulnerable to various security threats, since each platform server manages sensitive user data such as identities, passwords and biometric information [18]. For instance, in centralized metaverses, users only own their virtual avatars, while in decentralized metaverses, they can co-create virtual avatars in the DAO and collaborate with existing avatars to manage new ones. DAOs facilitate efficient operations, where any member, such as a holder or contractor, can propose initiatives. Proposals are discussed within the community and enacted if the majority votes in favor, with smart contracts recording the outcomes. Members who significantly contribute to the community gain reputation values, influencing future votes. Tokens circulate within the community to incentivize active participation.

TABLE I
Examples of IoE and MoE entities, key cybersecurity threats and countermeasures.

| IoE category | IoE examples | MoE (Digital Twins) | MoE examples | Cybersecurity threats | Cybersecurity countermeasures |
|---|---|---|---|---|---|
| People | Users connected to the Internet | Meta-users (Avatars) | Meta-users, company owners and marketing agencies | Avatar authentication, Identity theft, Data integrity, security and privacy (leakage in data transmission, data processing, in cloud/edge storage) | End-to-end advanced encryption, blockchain, post-quantum cryptography |
| Processes | Different types of processors: SBCs, Central Processing Units (CPUs), Road-Side Units (RSUs), Programmable Logic Controllers (PLCs), Supervisory Control and Data Acquisition (SCADA) Systems, Microcontrollers (MCUs) | Web 3.0 and Web 4.0 based components, AI agents | NFTs, Dynamic NFTs (dNFTs), DeFi, smart contracts and digital twins | Fraud, threats to digital asset ownership, threat to economic fairness, network-related attacks (e.g., DDoS, Sybil) | Blockchain, AI for threat detection, post-quantum cryptography |
| Data | Users' information, videos, music, documents, email | Generated data and DApps | Digital identity, data generated on MoE and related services (e.g., Intellectual Property) | Data tampering, false data injection, threats to ownership and provenance, vulnerabilities in data storage and access control | Blockchain for data management, post-quantum cryptography |
| Things | Sensors, vehicles, drones and AGVs (Autonomous Ground Vehicles), UAVs, robots, Radio Frequency Identification (RFID) tags, HVAC systems, cameras, smart irrigation systems, smart traffic lights | Virtual assets and services | Virtual worlds: virtual markets, virtual universities, virtual cities, virtual buildings, virtual environments | Fraud in virtual markets, exploits in IoT devices, rogue end devices, malware, unauthorized access | Blockchain, AI for threat detection, post-quantum cryptography |

Therefore, decentralization is essential, aiming to distribute control and decision-making away from centralized entities. Blockchain technology [15], with its transparent and immutable ledger, forms the backbone of decentralized systems. DApps and smart contracts enable trustless interactions, reducing the need for intermediaries and enhancing security.

Data management leverages Blockchain technologies to ensure data integrity, security and privacy. The provided key features include:

- Data are stored across a network of nodes, reducing the risk of data breaches and ensuring high availability.
- Advanced encryption techniques protect user data, ensuring privacy and compliance with regulations.
- Standardized protocols facilitate seamless data exchange between different systems and platforms within the MoE.

In the context of the Metaverse, decentralization offers several advantages, including increased user control over digital assets, enhanced privacy and resilience against attacks. However, it also presents challenges such as scalability, interoperability and regulatory compliance, which must be addressed to realize its full potential.

F. Opportunistic Edge Computing

OEC systems leverage Edge Computing devices to identify and provide computing services opportunistically to nearby IoT/IIoT devices and IoE entities. Edge Computing is used to eliminate the need for sending requests to a remote Cloud, so latency is significantly reduced [9]. Regarding IoT/IIoT devices and IoE entities, they are often resource-constrained (e.g., battery-powered) and distributed across extensive environments, and rely on external devices for connectivity and computational functions (e.g., processing power, internal storage), typically a remote Cloud that may not always be accessible. In addition, IoT/IIoT devices and IoE entities require reduced energy consumption for sustainability purposes.

OEC systems are particularly beneficial when IoT/IIoT devices and IoE entities lack continuous Internet connectivity or are static or have limited mobility, preventing them from relocating to communicate with other devices or entities [2].

Moreover, the collaborative nature of OEC solutions can help to mitigate certain restrictions of traditional Cloud Computing architectures, which are not designed for energy efficiency and face scalability issues with large-scale IoT/IIoT/IoE implementations [9].

Technological advancements in Single-Board Computers (SBCs), wearables, embedded IoT/IIoT devices and IoE entities have recently made OEC feasible and affordable. These devices now offer sufficient computing power and reduced power consumption, allowing them to function as MoE gateways or smart end devices within Edge Computing architectures.

The main software components of an OEC system for a MoE solution should include the following functionality:

- Peer discovery: it enables Meta-users to detect nearby IoT/IIoT devices and IoE entities and other Meta-users. This requires implementing low-latency secure device discovery protocols, as Meta-users usually move around, establishing communications between Metaverse applications and surrounding IoT/IIoT devices and IoE entities automatically.
- Peer routing: communications need to be routed to and from specific XR or IoT/IIoT devices and IoE entities, which require establishing an efficient path to the destination.
- Data routing: it allows for transmitting information from one device to another when the receiving device/entity is not within the communication range of the sending device/entity.
- Resource sharing: it is essential for optimizing resource use while delivering the necessary resources as close as possible to the IoT/IIoT/IoE and XR devices. Thus, efficient resource sharing allows for reducing response latency, which is crucial for the User Experience (UX) of MoE applications.

## III. Architecture of the MoE

### A. Main components

The architecture of the MoE consists of several core components that work together to create a cohesive and interactive digital environment. These components include:
- Digital Twin framework: it allows for creating and managing digital replicas of physical objects, enabling seamless integration between virtual and physical worlds.
- Blockchain infrastructure: it provides a decentralized ledger that ensures secure and transparent transactions within the MoE. It supports the use of smart contracts and digital assets.
- IoE integration layer: this layer connects IoT/IIoT devices and IoE entities among them, facilitating real-time data exchanges and interactions.
- User interfaces: they are the AR/MR/VR interfaces [2] through which users interact with the MoE, aiming to provide an immersive and intuitive user experience.

In practice, the previous components are deployed in layers like the ones illustrated in Figure 1. Such a Figure shows an example of MoE architecture for a Smart City and includes six main layers:
- Smart Object Layer: it is composed of the deployed smart IoT/IIoT devices and the IoE entities. Thus, this layer includes both objects from the real world (e.g., smart traffic lights, intelligent vehicles, devices for automating smart buildings) and virtual entities (e.g., digital twins of physical objects, digital messages, Artificial Intelligence (AI) agents, or virtual entities that represent data from processes or entered by people). This layer can exchange information not only with the MoE-ready XR Device Layer, but also with the servers of the Cloud Layer and with the Decentralized Subsystem Layer, which can feed/read data into/from the Smart Object Layer, and also interact with the deployed physical devices and the virtual entities. Thus, the smart objects of this layer track the state and changes of every IoT/IIoT surrounding object or IoE entity, sending notifications to the XR devices when needed, so that they can receive the latest information and events. Moreover, this layer also allows developers to exchange data that feeds the XR virtual content, which can react to changes in real-world IoT/IIoT devices and IoE entities. Furthermore, this layer can send messages back to the real devices in response to the user's interactions (like clicks or when grabbing a specific virtual object).
- MoE-ready XR Device Layer: it includes all the XR devices used by the Meta-users to interact with the Smart Object Layer. Every XR subsystem runs locally the necessary software to both control user interactions (e.g., hand movements, eye tracking) and communications with IoT/IIoT devices and IoE entities. Moreover, to keep synchronized with the virtual elements of a MoE world, MoE-ready XR devices should be able to monitor the state of the components shared with other Meta-users. Furthermore, XR devices have three other communication channels:
    - XR device-to-XR device (XR2XR): two MoE-ready XR devices can communicate directly between them or by using OEC communications, thus avoiding unnecessary connections through external devices.
    - XR device-to-OEC device (XR2OEC): in case of having an OEC device in range, XR devices can use it to exchange data with third parties (e.g., with other XR devices or Smart Objects that are not in range, with the Cloud or with a Decentralized Subsystem) or to perform certain processing tasks that require a high amount of computing power.
    - XR device-to-Cloud (XR2Cloud) and XR device-to-Decentralized System (XR2DS): type of communications that makes use of the Routing Layer to exchange data either with a remote Cloud certain services of a Metaverse are provided or with specific decentralized services (e.g., with a blockchain or with a decentralized storage).
- OEC Layer: it provides opportunistic Edge Computing services to the MoE devices whenever they are in range. In Figure 1, as an example, three types of OEC devices are included: a low-power computing device (based on the use of a SBC), a high-power computing device (a Cloudlet based on a powerful server) and a mobile device (a smartphone).
- Routing Layer: this layer includes a set of gateways that allow for exchanging requests and data between the Cloud Layer and the Decentralized Subsystem Layer with the deployed XR and OEC devices.
- Cloud Layer: it is where the actual MoE applications are executed. Thus, it allows the different metaverses to provide service for diverse fields, like those illustrated in Figure 1.
- Decentralized Subsystem Layer. They enable the

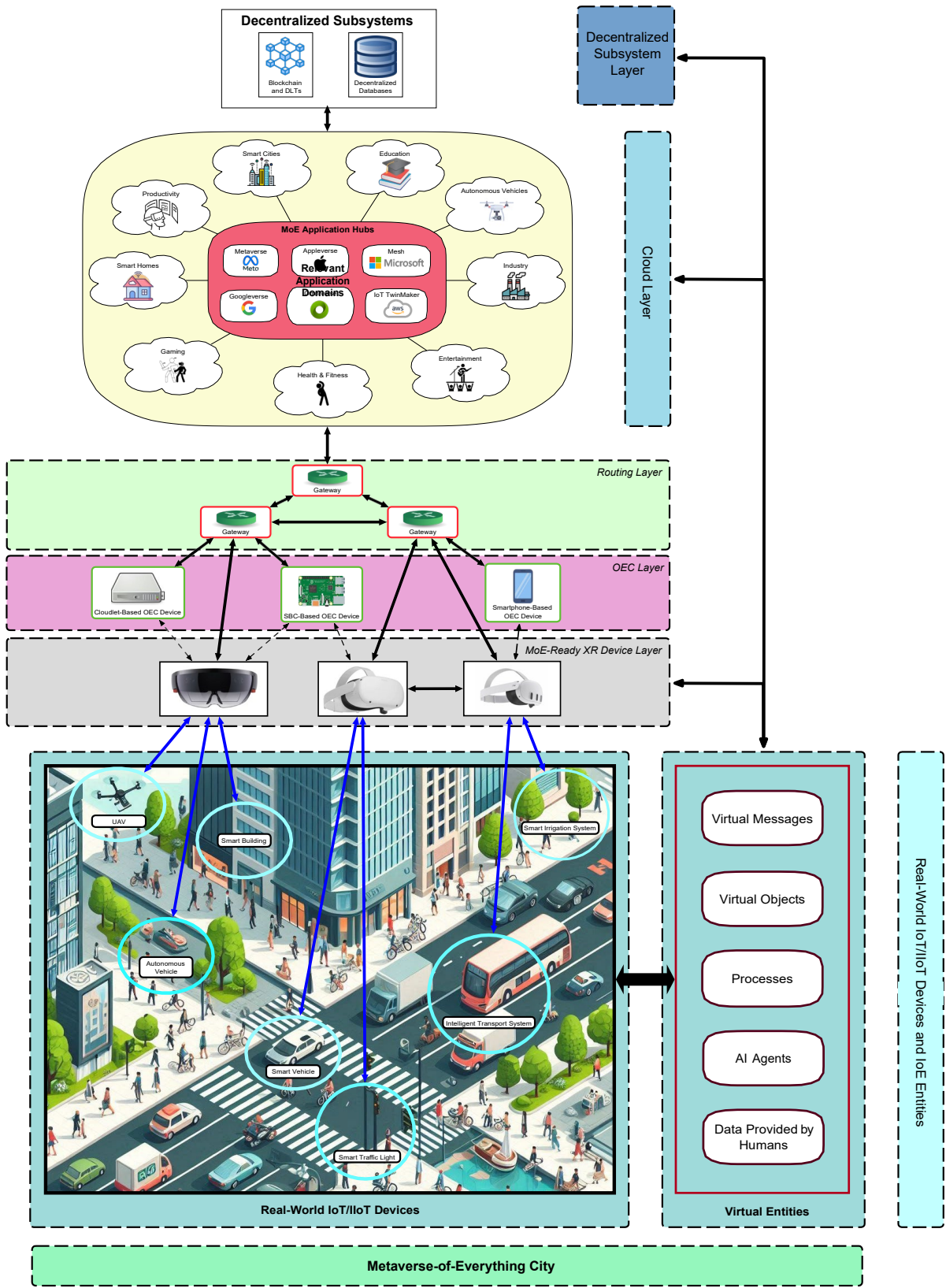

Fig. 1. Example of communications architecture for a MoE city.

decentralization of the MoE by providing services related to blockchain, DLTs or decentralized storage (e.g., decentralized databases that utilize InterPlanetary File System (IPFS)). For instance, DAOs and smart contracts facilitate efficient data transmission and verification within the Metaverse's economic system. Consensus mechanisms can guarantee transactions, while distributed blockchain storage enhances security by providing redundancy. Further networking details in blockchain-enabled communications and computing can be found in [19].

## IV. Challenges and recommendations

### A. Meta-user interaction and UX

The MoE promises a world of incredible possibilities, yet for its full potential to be realized, current IoT/IIoT/IoE technologies need significant advancement and further research is necessary on the optimization of the OEC layer, considering metrics related to throughput, task queuing mechanisms or failure recovery protocols. This need for progress to achieve truly immersive experiences hinges on four crucial elements of user experience:

- Seamless integration: a fluid interplay between physical and virtual environments is essential. This involves developing robust interfaces and data-sharing mechanisms that harmonize real-world sensor data with XR content.
- Synchronization: real-time experiences must be consistently aligned for all participants, necessitating ultra-low-latency connections. IoT, IIoT and IoE demand an ideal end-to-end latency below 10 ms to enable synchronous interactions within XR environments.
- Rapid interaction: users should be able to swiftly engage with both digital content and physical surroundings. This includes supporting instantaneous hand, body and object tracking, as well as haptic feedback and other sensory modalities that further immerse Meta-users in virtual scenarios.
- Persistence: actions and states across XR and real-world interactions must be retained, ensuring that user activities, system states and environmental data remain consistent over time.

### B. Energy efficiency

Given that many IoT/IIoT devices are battery-operated and have limited processing power, the increased data exchange with XR devices can quickly deplete energy resources. This challenge is compounded by the demands of real-time communication and high-fidelity rendering in the MoE. To address these issues it is necessary:

- To design optimized architectures: efficient MoE architectures must incorporate strategies such as intelligent offloading (e.g., to edge servers) or sleep scheduling to minimize energy consumption.
- To optimize blockchains and DLTs: specific techniques to reduce the computational overhead of blockchain and DLTs [15] (particularly in consensus and verification processes) are essential for extending IoT/IIoT device battery life.
- To devise Green Metaverses: designing for sustainability from the outset promotes greener solutions [6], ensuring that the MoE can scale responsibly and remain environmentally sustainable.

### C. Diversity of real-world scenarios

The MoE must be adaptable to a wide range of application domains, each with unique constraints and requirements. Examples of such application domains are:

- Healthcare: privacy, safety and reliability are critical, especially where patient monitoring and telemedicine services intersect with virtual elements.
- Education: applications like immersive classrooms demand solutions that are affordable, user-friendly and robust to changes in the environment.
- Transportation and Smart Cities: real-time data from vehicles and infrastructure must integrate seamlessly with virtual overlays, enabling safer navigation and more efficient public services.
- Industrial environments: ultra-low latency and high reliability are vital for controlling machinery or monitoring production lines, necessitating stringent performance and safety guarantees.

### D. Open-source opportunistic plug-and-play protocols

The MoE should be able to make use of a wide variety of connected devices and data sources, which requires protocols that support opportunistic discovery, rapid secure communication exchanges. Thus, future researchers should contribute to:

- Open-source frameworks: the development of community-driven standards and software fosters faster innovation and broader interoperability among XR devices, IoT/IIoT devices and IoE entities.
- Seamless integration: plug-and-play approaches ensure that Single-Board Computers (SBCs), sensors, actuators and other IoT/IIoT elements can be quickly configured and integrated into Metaverse applications without tedious manual setups.
- Security and trust: as new devices join or leave the network opportunistically, robust authentication and authorization mechanisms are crucial for maintaining a secure trustworthy environment.

### E. Advanced communication and networking

Research on Ultra-Low Latency Networks (ULLNs) focuses on communication technologies (e.g., 5G/6G, WiFi 7/8) able to provide latencies of less than 10 ms and an optimal trade-off in terms of high throughput and very low round-trip delay (1-10 ms). Latencies can range from 0.25–10 ms for data sizes of 10–300 bytes in industrial automation applications, 1 ms for the Tactile Internet with data sizes of 250 bytes, or 0.4–2 ms for 12–16 kB in AR/VR applications (all with a reliability of at least 99.999%). Locally-deployed Edge Computing devices enable the projection of large virtual objects with relatively low latency, typically averaging around 250 ms. However, latency should be further reduced when integrating data from IoT/IIoT devices, as certain sensors and actuators operate at higher update rates [2]. The optimal latency is achieved by jointly optimizing various factors such as edge caching, communication protocols, task offloading, multi-tier computing, bandwidth allocation, or transmit power. Efficient resource allocation is also crucial for achieving optimal performance in the Metaverse [12]. Furthermore, Software-Defined Networking (SDN) enables the creation and independent management of virtual networks, thereby enhancing network efficiency and flexibility.

F. Standardization

Standardization is an ongoing effort, so it is necessary to monitor the different organizations (e.g., IEEE, ITU's Focus Group on Metaverse, 3GPP's Metaverse projects, IETF and MPAI) that are carrying out initiatives for such a purpose [20].

## V. Conclusions

The journey towards a Blockchain and Opportunistic MoE represents a significant leap in the evolution of the Internet. This ambitious vision integrates Blockchain with Opportunistic Edge Computing to create a robust interconnected network that seamlessly blends physical and digital worlds. The MoE aims to enhance security, efficiency and scalability, paving the way for innovative applications and transformative user experiences across various industries. The article proposed a comprehensive vision the main concepts of the MoE, described its communications architecture and shed light on the integration of XR, blockchain, IoE and opportunistic communications, providing valuable insights into the field.

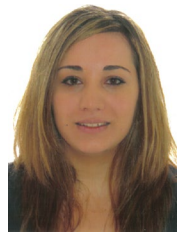

Paula Fraga-Lamas has been a researcher at University of A Coruña (UDC) since 2009. She has more than 100 publications and has participated in more than 40 research projects funded by the regional and national government as well as R&D contracts. Her current research interests include IoT/IIoT/IoE, Cyber-Physical Systems, Extended Reality (XR), fog and edge computing, blockchain and Distributed Ledger Technology (DLT), and cyber-security. She is a Senior Member of the IEEE. Contact her at paula.fraga@udc.es.

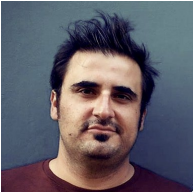

Sérgio Ivan Lopes is the Director-General at CiTin – Centro de Interface Tecnológico e Industrial, Assistant Professor at the Technology and Management School of the Polytechnic Institute of Viana do Castelo (ESTG-IPVC), and researcher with the Instituto de Telcomunicações. His current research interests include Cyber-Physical Systems, IoT/IIoT, Industry 5.0, Sensor Systems, and Edge Intelligence. In such areas, he has co-authored more than 100 scientific publications. He is a Senior Member of the IEEE. Contact him at sil@estg.ipvc.pt.

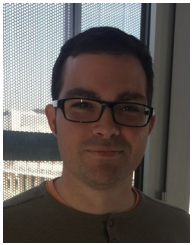

Tiago M. Fernández-Caramés works as an Associate Professor at University of A Coruña (UDC) (Spain). His current research interests include XR technologies, IoT/IIoT systems and blockchain, as well as the other different technologies involved in the Industry 4.0 and 5.0 paradigms. In such fields, he has contributed to more than 120 publications. He is a Senior Member of the IEEE. Contact him at tiago.fernandez@udc.es.